\documentclass[12pt,preprint]{aastex}

\usepackage{natbib}
\usepackage{color}

\shorttitle{1D Radiative Transfer Intercomparison}
\shortauthors{Yang et al.}

\begin{document}



\title{Differences in water vapor radiative transfer among 1D models
  can significantly affect the inner edge of the habitable zone}

\author{Jun Yang$^{1, 2}$, J\'{e}r\'{e}my Leconte$^3$, Eric T. Wolf$^4$,
  Colin Goldblatt$^5$, Nicole Feldl$^6$, Timothy Merlis$^7$, Yuwei
  Wang$^1$, Daniel D.B. Koll$^8$, Feng Ding$^8$, Fran\c{c}ois
  Forget$^3$, and Dorian S. Abbot$^8$}
\affil{$^1$Department of Atmospheric and Oceanic Sciences, School of Physics, Peking University, Beijing, China\\
  $^2$Previously at Department of the Geophysical Sciences, University of Chicago, Chicago, IL, USA\\
  $^3$Laboratoire de M\'et\'eorologie Dynamique, Institut Pierre Simon Laplace, CNRS, Paris, France\\
  $^4$Laboratory for Atmospheric and Space Physics, University of Colorado in Boulder, Colorado, USA \\
  $^5$School of Earth and Ocean Sciences, University of Victoria, Victoria, BC, Canada \\
  $^6$Division of Geological and Planetary Sciences, California Institute of Technology, CA, USA\\
  $^7$Department of Atmospheric and Oceanic Sciences at McGill University, Montr\'{e}al, Canada\\
  $^8$Department of the Geophysical Sciences, University of Chicago, Chicago, IL, USA}
\email{Correspondence:~junyang@pku.edu.cn, abbot@uchicago.edu}

\begin{abstract}
  An accurate estimate of the inner edge of the habitable zone is
  critical for determining which exoplanets are potentially habitable
  and for designing future telescopes to observe them. Here, we
  explore differences in estimating the inner edge among seven 
  one-dimensional (1D) radiative transfer models: two line-by-line
  codes (SMART and LBLRTM) as well as five band codes (CAM3,
  CAM4\_Wolf, LMDG, SBDART, and AM2) that are currently being used in
  global climate models. We compare radiative fluxes and spectra in
  clear-sky conditions around G- and M-stars, with fixed moist
  adiabatic profiles for surface temperatures from 250 to 360
  K. We find that divergences among the models arise mainly from 
  large uncertainties in water vapor absorption in the window region
  (10 $\mu$m) and in the region between 0.2 and 1.5 $\mu$m.
  Differences in outgoing longwave radiation increase with surface
  temperature and reach 10-20~W\,m$^{-2}$; differences in shortwave
  reach up to 60 W\,m$^{-2}$, especially at the surface and in the
  troposphere, and are larger for an M-dwarf spectrum than a solar
  spectrum. Differences between the two line-by-line models are
  significant, although smaller than among the band models. Our
  results imply that the uncertainty in estimating the insolation threshold 
  of the inner edge (the runaway greenhouse limit) due only to 
  clear-sky radiative transfer is $\approx$\,10\,\% of modern Earth's solar 
  constant (i.e., $\approx$\,34 W\,m$^{-2}$ in global mean) among 
  band models and $\approx$\,3\,\% between the two line-by-line models. 
  These comparisons show that future work is needed focusing on improving 
  water vapor absorption coefficients in both shortwave and longwave, as 
  well as on increasing the resolution of stellar spectra in broadband models.

\end{abstract}

\keywords{astrobiology --- planets and satellites: atmospheres 
---  planets and satellites: general --- methods: numerical --- radiative transfer}








\section{Introduction}
About 1600 planets orbiting other stars have been confirmed, and their
number is constantly rising. A critical question is which of these
planets are potentially habitable. Because liquid water is necessary
for all known life on Earth, potentially habitable planets are
generally defined as planets in the region around a star where they
can maintain liquid water on their surface \citep{Kasting2010}. This
region is called the habitable zone \citep{Kastingetal1993,
  Kastingetal2014}. A standard assumption for the habitable zone is
that the atmosphere is mainly composed of H$_2$O, CO$_2$, and N$_2$.
Another critical assumption is that the silicate-weathering feedback
functions on exoplanets. This feedback has been proposed based on
studies of Earth's history and would regulate CO$_2$ through the
dependence of CO$_2$ removal by silicate weathering on planetary
surface temperature and precipitation \citep{Walkeretal1981}. If the
the silicate-weathering feedback is functioning, the atmospheric
CO$_2$ concentration should be low for planets near the inner edge of
the habitable zone and high for planets near the outer edge of the
habitable zone.

In this study, we focus on the inner edge of the habitable zone, which
is determined by the loss of surface liquid water through
either a moist greenhouse state or a runaway greenhouse state
\citep{Kasting1988, Nakajimaetal1992, Abe1993}. The
  purpose of this study is to calculate the inner edge of the
  habitable zone (in particular the runaway greenhouse limit) using
  seven 1D, cloud-free radiative transfer models, and to compare the
  differences among the models.

For both moist and runaway greenhouse states, the key process is the
water vapor feedback: As the stellar flux increases, surface and
atmospheric temperatures increase; the saturation water vapor pressure
increases approximately exponentially with temperature following the
Clausius--Clapeyron relation; and the increased atmospheric water
vapor further warms the surface and the atmosphere because water vapor
is a strong greenhouse gas and a good shortwave absorber. Once the
mixing ratio of water vapor in the stratosphere becomes very high, the
loss of water to space via photolysis and hydrogen escape will be
significant. This process is called the moist greenhouse. For
instance, if the stratospheric water vapor volume mixing ratio exceeds
$\approx$\,3$\times$10$^{-3}$, Earth would lose an entire ocean's
worth of water (1.3\,$\times$\,10$^{18}$\,m$^{3}$) over the age of the
Solar system \citep[$\approx$\,4.6 billion years,][]{Kasting1988}.

The runaway greenhouse state arises from an energy imbalance, in which
the atmosphere becomes optically thick in all infrared wavelengths,
and absorbed stellar flux exceeds thermal infrared emission
\citep{Pierrehumbert2010, GoldblattandWatson2012}. In the runaway
greenhouse, a planet will keep warming until all surface liquid water
evaporates. During a runaway greenhouse the stratosphere typically
becomes moist and water escapes rapidly to space. A third, empirical
limit of the inner edge of the habitable zone has also been defined
based on the fact that Venus may have had liquid water on its surface
until about one billion years ago \citep{Kastingetal1993}, although
this limit may depend on planetary rotation rate \citep{Yangetal2014}.

The inner edge of the habitable zone was estimated using a 1D climate
model by \citet{Kastingetal1993} and
\citet{Abe1993}, and recently this work was updated by
\citet{Kopparapuetal2013, Kopparapuetal2014} and
\citet{Kastingetal2014}. Using a 1D model and assuming a cloud-free
and saturated atmosphere, \citet{Kopparapuetal2014} computed the inner
edge of the habitable zone around the present Sun to be 0.99, 0.97,
and 0.75 AU for the moist greenhouse limit, the runaway greenhouse
limit, and the recent Venus limit, respectively. The corresponding
solar fluxes are 1380, 1420, and 2414~W\,m$^{-2}$, respectively, which
can be compared to Earth's present-day solar flux of about
1360~W\,m$^{-2}$. For K and M stars, the insolation threshold of the
inner edge is smaller due to the fact that a redder stellar spectrum
causes a lower planetary albedo. Meanwhile, K and M stars are cooler
and smaller than the Sun, so that the inner edge is located much
closer to the host star. For F stars, the conditions are opposite to
those of K and M stars, so that the insolation threshold is larger,
and the habitable zone is farther away from the host star
\citep{Kastingetal1993}. Apart from the stellar spectrum, other
factors can also influence the insolation threshold, for example
planetary gravity \citep{Pierrehumbert2010, Kopparapuetal2014}
and background gas concentrations (such as N$_2$;
  \cite{Goldblattetal2013}). Furthermore, the inner edge of
  the habitable zone of dry planets (desert worlds with limited
  surface water) may be much closer to the host stars
  \citep{Abeetal2011, Kodamaetal2015}.

Besides the work of \citet{Kastingetal1993} and
\citet{Kopparapuetal2013}, other 1D radiative transfer models have
also been employed to estimate the inner edge of the habitable zone.
The inter-model differences are significant and the predicted
insolation thresholds for the runaway greenhouse limit vary
by up to $\approx$\,80~W\,m$^{-2}$ (i.e., 20~W\,m$^{-2}$ of radiation
impinging on the planet after geometric factors are accounted for).
Using SMART, \citet{Goldblattetal2013} found that the insolation
threshold of the inner edge for a pure vapor atmosphere is
$\approx$\,1340 W\,m$^{-2}$. In contrast, \citet{Kopparapuetal2013}
and \citet{Leconteetal2013} found a threshold of $\approx$\,1420
W\,m$^{-2}$ in Kasting's radiative transfer model and in the 1D model
LMDG, even though all groups used the same surface albedo of 0.25 and
a pure water vapor atmosphere. This inter-model spread in the
position of the inner edge represents almost 10\% of the total flux
difference between the inner and outer edge, which is
$\approx$910--960~W\,m$^{-2}$ \citep{Kopparapuetal2013}. We note
that the above papers differed primarily in their treatment of
radiative transfer, whereas additional physical processes, such as
atmospheric dynamics and clouds
\citep{Leconteetal2013,Yangetal2013,WolfandToon2014,Yangetal2014,WolfandToon2015},
could lead to even larger discrepancies in estimates of the habitable
zone.

To explore the physical processes that determine the inner edge of the
habitable zone, we therefore organized an exoplanet climate model
inter-comparison program. In this paper we focus entirely on clear-sky
radiative transfer and explore the differences among seven 1D
radiative transfer models with the same temperature, humidity, and
atmospheric compositions. We find that the uncertainty in radiative
transfer leads to $\approx$\,10\,\% variations in estimates of the
inner edge of the habitable zone and identify the radiative transfer
of water vapor as the main culprit for discrepancies among models. We
have organized the paper as follows: Section 2 describes the models we
use and the experimental design. Section 3 presents the longwave
results and Section 4 presents the shortwave results. We discuss our
results in Section 5 and conclude in Section 6.


\section{1D Radiative Transfer Model Intercomparison}

The pure radiative-transfer models we use are SBDART, 1D versions of
CAM3, CAM4\_Wolf, AM2, and LMDG, and two line-by-line radiative
transfer models, SMART and LBLRTM. The main properties of the models
are summarized in Table~\ref{model-descriptions}. Given temperature
and water vapor profiles, these models calculate radiative fluxes at
each vertical level. They employ different spectral line databases,
different radiative transfer schemes, different H$_2$O continuum
absorption methods, and different multiple scattering schemes.

\begin{figure}
\begin{center}
\includegraphics[angle=0, width=140mm]{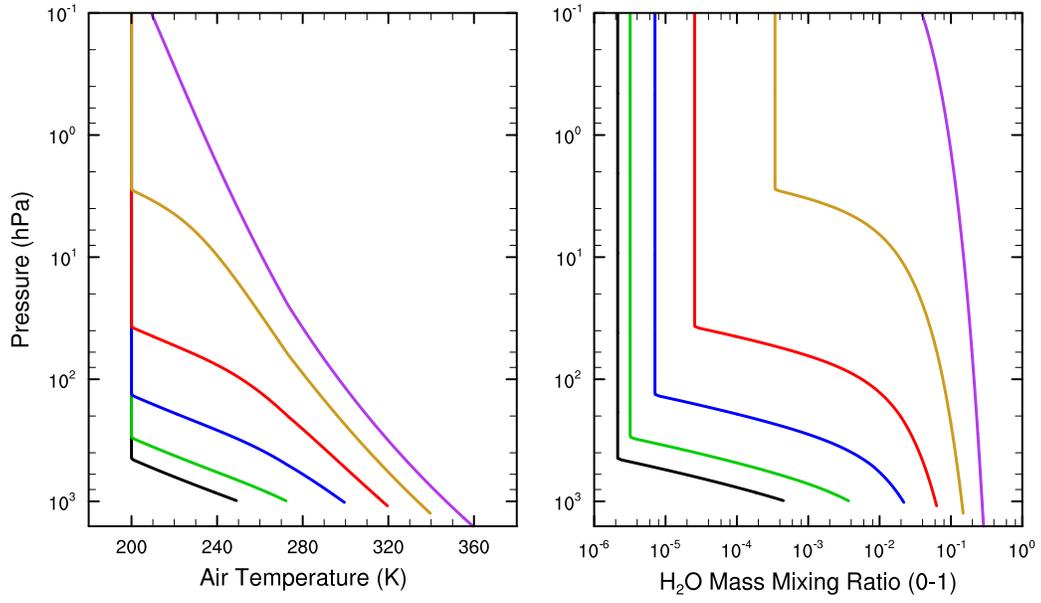}
\vspace{-30mm}
\caption{Input (a) air temperature and (b) water vapor mass mixing
  ratio used in the radiative-transfer calculations. Surface
  temperatures are 250, 273, 300, 320, 340, and 360 K.}
\label{fig-1D-T-Q-profiles}
\end{center}
\end{figure}

Our calculations are performed in a similar manner to
\citet{Kasting1988} and \citet{Kastingetal1993}, but cover a smaller
range of temperatures, as shown in Fig.~\ref{fig-1D-T-Q-profiles}.
Specifically, we perform all shortwave calculations with a
  top-of-atmosphere fixed stellar flux (340\,W\,m$^{-2}$) and longwave
  calculations with a variety of fixed surface temperatures and
  assumed atmospheric temperature profiles. Note that the outgoing
  longwave radiation (infrared emission to space) from our longwave
  calculations does not necessarily balance the absorbed shortwave. In
  order to interpret our results, we will make use of the ``effective
  solar flux'' concept \cite{Kasting1988} in section 5 to estimate the
  solar flux that would allow equilibrium, which we will explain at
  that point.

We set the surface temperature to 250, 273, 300, 320, 340, and
360~K\footnote{We did not examine temperatures higher than 360 K,
  because \cite{Leconteetal2013} have shown that Earth may enter into
  the runaway greenhouse state when globally averaged surface
  temperature is $\approx$\,340~K.}. We were not able to get the AM2
scheme to converge for temperatures of 320~K and above, so we will
only address the results of AM2 experiments at 250, 273, and 300 K.
The temperature structures are moist adiabatic profiles overlain by a
200-K isothermal stratosphere. The atmosphere is assumed to be
saturated in water vapor (relative humidity is equal to one). The
volume mixing ratio of water vapor in the stratosphere is set equal to
its value at the tropopause. The atmosphere is assumed to be
Earth-like, namely 1-bar N$_2$, variable H$_2$O, and 376~ppmv CO$_2$.
We do not include other gases, clouds, or aerosols. Because the total
pressure of the atmosphere is variable as temperature changes
and because we fix the volume mixing ratio of CO$_2$, the
absolute mass of CO$_2$ is not constant, but this should not affect
our results significantly for CO$_2$ at such low concentrations.
For example, as the surface temperature increases from 300 to
  360 K, the surface pressure increases from 1019 to 1598 hPa due to
  the increase water vapor, and the vertically integrated mass of
  376-ppmv CO$_2$ increases from 5.7 to 8.9 kg\,m$^{-2}$. If assume
  the radiative forcing at the tropopause for doubling CO$_2$
  concentration is about 4 W\,m$^{-2}$ \citep{Collinsetal2006}, the
  increased radiative forcing due to the increase of CO$_2$ mass is
  about 2.1 W\,m$^{-2}$. This forcing should be uniform across models
  (so it will not cause discrepancies) and it is one order of
  magnitude smaller than the differences of tens of W\,m$^{-2}$ in
  radiative fluxes at high temperatures among the models found in our
  calculations below.

The incoming stellar flux at the top of the models is 340\,W\,m$^{-2}$
in all calculations. By default, the surface has a uniform albedo of
0.25 in the shortwave. An exception is the surface albedo in SMART,
which was set to 0.25 for wavelengths shorter than 3 $\mu$m and to
zero at longer wavelengths, due to a mistake on our part. This means
SMART absorbs more shortwave radiation at the surface and
underestimates the reflected, upward shortwave flux. The magnitude of
this underestimation, however, should be less than
$\approx$\,1.6~W\,m$^{-2}$, due to the fact that only a small fraction
of the stellar energy is in the region of wavelengths longer than 3
$\mu$m for both G and M stars. For longwave calculations, the surface
is assumed to have a uniform emissivity equal to one. All models have
a top pressure of 0.1~hPa. The number of vertical levels is 75 in
SMART, 150 for longwave and 75 for shortwave in LBLRTM, and 301 in the
band models. Due to the high spectral resolution and the
  long-time integrations, we had to limit the number of vertical
levels in the two line-by-line models (SMART and LBLRTM) so that the
calculations would be numerically feasible. The vertical resolution
has a very small effect on radiative fluxes, less than
$\approx$\,0.01~W\,m$^{-2}$ \citep{Collinsetal2006}. The solar zenith
angle is 60$^{\circ}$ in all the calculations.

\begin{figure}
\begin{center}
\includegraphics[angle=0, width=170mm]{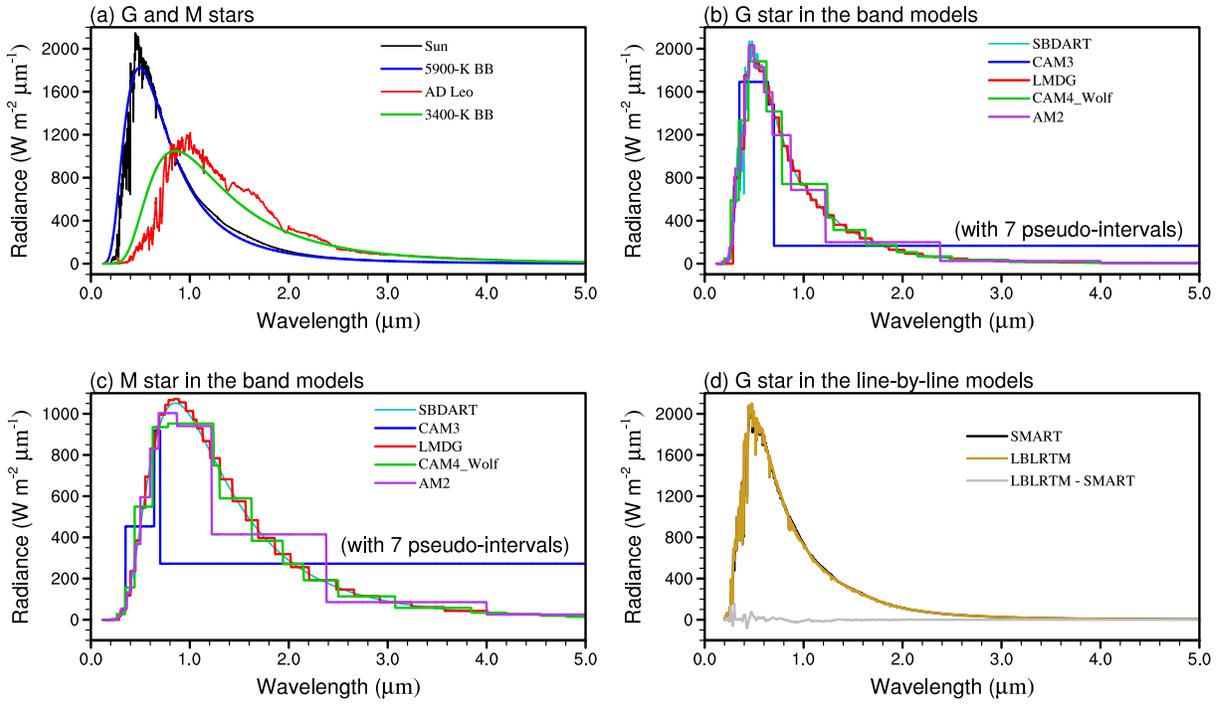}
\vspace{-50mm}
\caption{(a) Realistic stellar spectra and the corresponding blackbody
  spectra. (b) Representation of the G-star spectrum in the band
  models. (c) Same as (b), but for an M star. There are 7  
  pseudo-intervals at the near-infrared region in CAM3; in AM2, there
  are 38 pseudo-intervals for all bands, which are not shown in the figure.
  (d) Representation of the G-star spectrum in two line-by-line
  models and the difference between them. For comparison, both SMART and LBLRTM in (d) have been
  converted to have a spectral resolution of approximately 0.0025
  $\mu$m. In all panels, differences at wavelengths longer than 5 $\mu$m are very
  small and thus are not shown.}
\label{fig-four-spectrums}
\end{center}
\end{figure}

We explore two stellar spectra, the real solar spectrum and an
idealized, 3400-K blackbody spectrum (representing an M dwarf),
except where explicitly noted otherwise. These two spectra,
as well as other two spectra (a 5900-K blackbody and the real spectrum
of the M dwarf AD Leo), and their representations in the models are
shown in Fig.~\ref{fig-four-spectrums}. The wavelength corresponding
to the maximum spectral radiance in units of
W\,m$^{-2}$\,$\mu$m$^{-1}$ is $\approx$\,0.5 $\mu$m for the G star and
$\approx$\,0.8 $\mu$m for the M star. The two line-by-line models have
hundreds of thousands of spectral intervals. SBDART is a
high-resolution band model and has 369 intervals in the stellar
spectrum. In the broadband models, the stellar spectrum is divided
into 19, 18, 23, and 36 spectral and pseudo-spectral intervals in
CAM3, AM2, CAM4\_Wolf, and LMDG\footnote{The LMDG version used in
  Leconte et al. (2013) has only 19 bands in the stellar spectrum.
  Thus, 1D LMDG results from this study may not be strictly relevant
  to the results of Leconte et al. (2013).}, respectively. For
instance, CAM3 has seven spectral intervals for O$_3$, one for the
visible, three for CO$_2$, and seven near-infrared, pseudo-spectral
intervals for H$_2$O. These pseudo-spectral intervals are employed to
keep the number of spectral intervals as small as possible while
fitting radiative heating rates to be close to the results of
line-by-line calculations \citep{Briegleb1992}. The spectral intervals
are finer in the visible region than in the near-infrared region in
all band models. This is because these models were developed to
simulate the climates of planets with Earth-like atmospheres and with
Sun-like host stars. LMDG uses 16 gauss points for calculations of the
cumulated distribution function of absorption data for each spectral
interval, and CAM4\_Wolf uses 8 gauss points per interval except at
the intervals between 500 and 820 cm$^{-1}$, where 16 gauss points are
used.

The output quantities from each model include (1) shortwave and
longwave fluxes at the surface and at the top of the atmosphere, (2)
upward and downward shortwave and longwave fluxes at each level of the
atmosphere, (3) (optionally) spectra at the surface and/or at the top
of the atmosphere. In this paper, for convenience 
and consistent with standard terminology, ``longwave'' refers to the thermal 
infrared emission from the planet, and ``shortwave'' refers to the stellar
energy from the star. Although the shortwave and longwave overlap
somewhat in the near-infrared, treating them separately still
conserves energy, since the shortwave module does not consider the
thermal energy and the longwave module does not consider the
stellar energy.


\section{Comparison of Longwave Radiation}\label{section-results-LW}

\begin{figure}
\begin{center}
\includegraphics[angle=0, width=140mm]{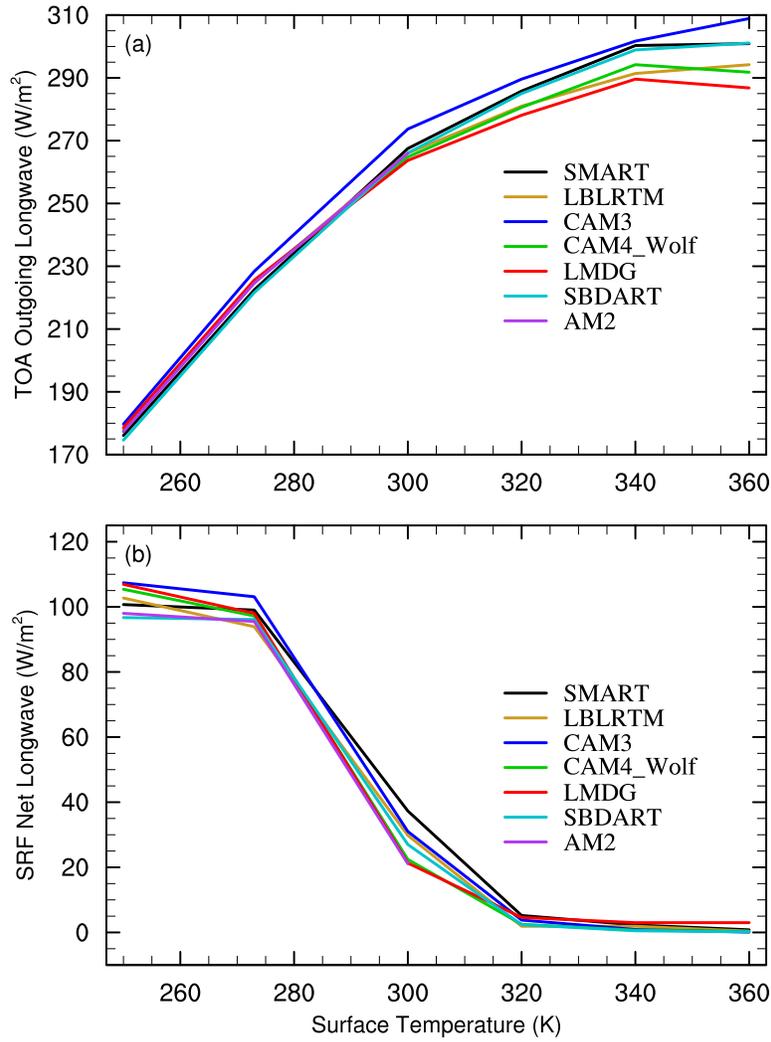}
\vspace{0mm}
\caption{(a) Outgoing longwave radiation at the top of the atmosphere
  (TOA), and (b) net longwave radiation at the surface (SRF) as a
  function of surface temperature for all of the models from 250 to
  360~K.}
\label{fig-OLR-1D}
\end{center}
\end{figure}

\begin{figure}
\begin{center}
\includegraphics[angle=0, width=165mm]{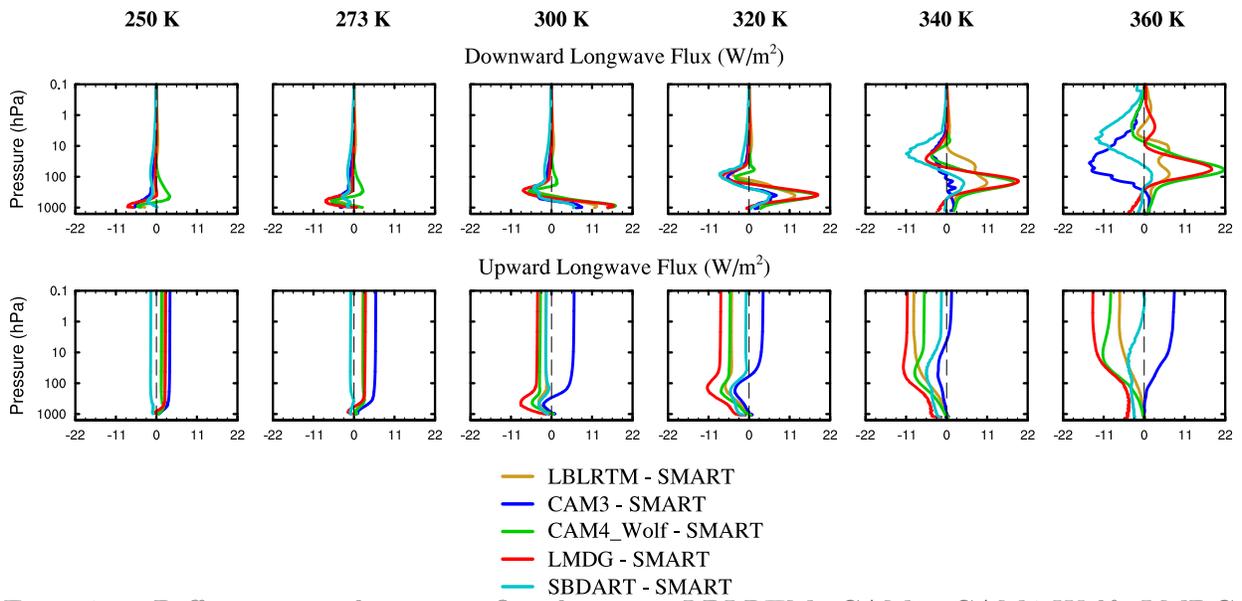}
\vspace{-50mm}
\caption{Differences in longwave flux between LBLRTM, CAM3,
  CAM4\_Wolf, LMDG, SBDART, and SMART as a function of pressure and
  for surface temperatures from 250 to 360~K. Upper panels: differences in downward
  longwave flux; lower panels: differences in upward longwave flux. Note that longwave Fluxes from AM2 are unavailable.}
\label{fig-flux-pressure-12panels}
\end{center}
\end{figure}

Outgoing longwave radiation (OLR) fluxes at the top of the atmosphere
(TOA) as a function of surface temperature are shown in
Fig.~\ref{fig-OLR-1D}(a). At low temperatures, all the models agree
well with each other, while at high temperatures, the differences
among the models become larger. The model spreads in the OLR are 
5, 10, 17, and 25~W\,m$^{-2}$ at surface temperatures of 250, 
300, 320, and 360~K, respectively. At the surface, the differences in net
longwave flux are relatively small, less than 15~W\,m$^{-2}$
(Fig.~\ref{fig-OLR-1D}(b)), since the atmosphere near the surface
becomes optically thick, especially once the surface temperature is
above 320 K. In the troposphere, the differences in downward longwave
flux can be greater than those at the surface and at the TOA
(Fig.~\ref{fig-flux-pressure-12panels}).

In general, LMDG has the lowest OLR and the strongest greenhouse
effect, whereas CAM3 has the highest OLR and the weakest greenhouse
effect. In LMDG, SMART, and CAM4\_Wolf, the OLR curves level out as
the surface temperature is increased from 340 to 360~K. In CAM3,
SBDART, and LBLRTM, however, the OLR curves keep increasing although
at a small rate. For these three models, the maximum surface
temperature of 360~K examined here may be not high enough to obtain
their OLR limit, or they do not become optically thick at all
wavelengths at high temperatures \citep{Goldblattetal2013}. The existence 
and the value of the OLR limit are very
important for the runaway greenhouse state. In LMDG, SMART, 
and CAM4\_Wolf, the OLR limit is $\approx$\,287, 301, and 292 W\,m$^{-2}$, respectively.

\begin{figure}
\begin{center}
\includegraphics[angle=0, width=165mm]{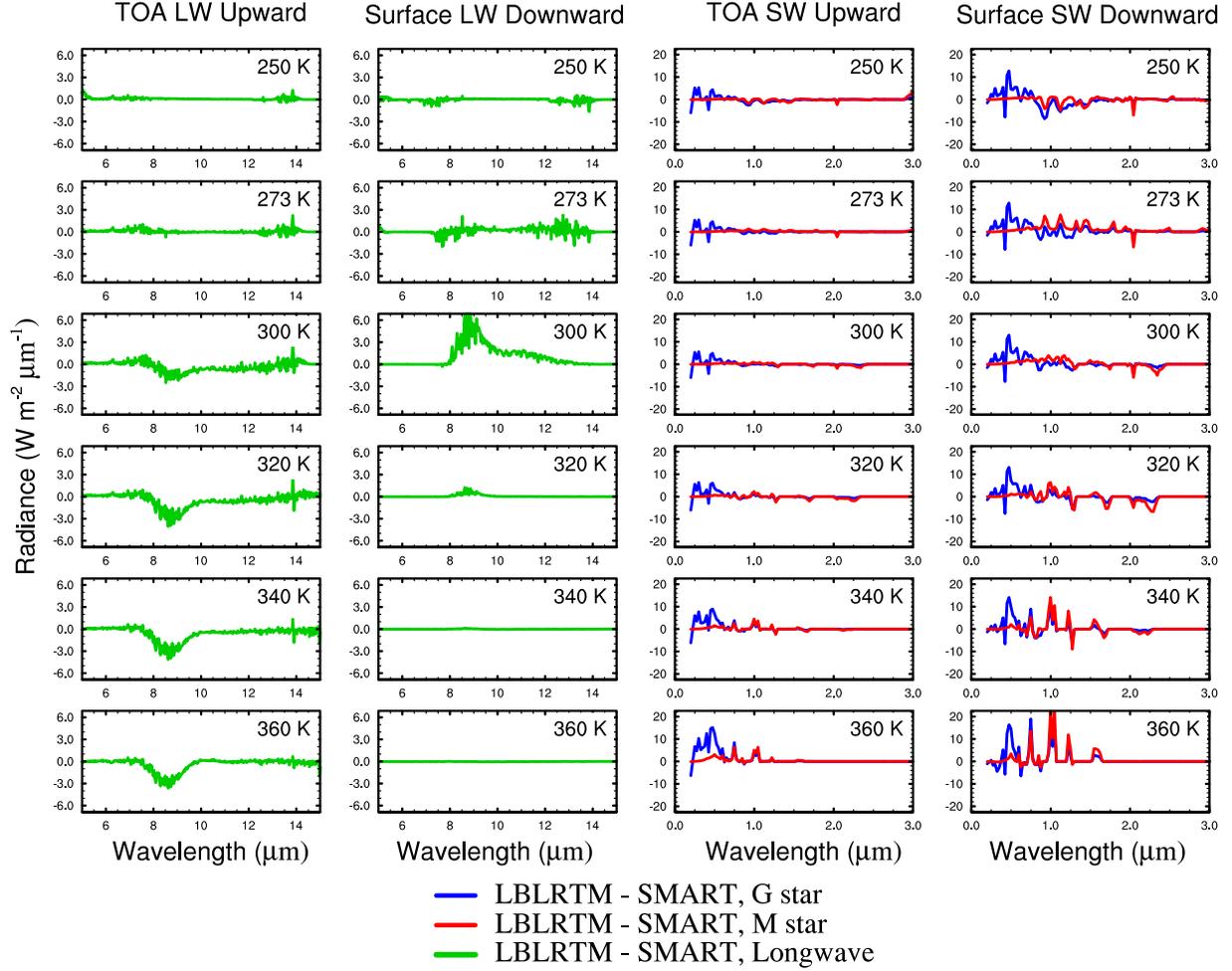}
\vspace{-40mm}
\caption{Shortwave (SW) and longwave (LW) spectral differences between
  the two line-by-line models, LBLRTM minus SMART. First column:
  upward longwave at the top of atmosphere; second column: downward
  longwave at the surface; third column: upward shortwave at the top
  of atmosphere; and fourth column: downward shortwave at the surface.
  For comparison, both models have been converted to have a spectral
  resolution of approximately 0.0025 $\mu$m. Differences at wavelengths
  between 3 and 5 $\mu$m and longer than 15 $\mu$m are relatively
  small and thus are not shown. In the third column, a part of the
  difference in wavelengths shorter than 1.0 $\mu$m is due to the
  difference in input solar spectra (see Fig.~2(d)).}
\label{fig-colin-LW-spectra}
\end{center}
\end{figure}

\begin{figure}
\begin{center}
\includegraphics[angle=0, width=165mm]{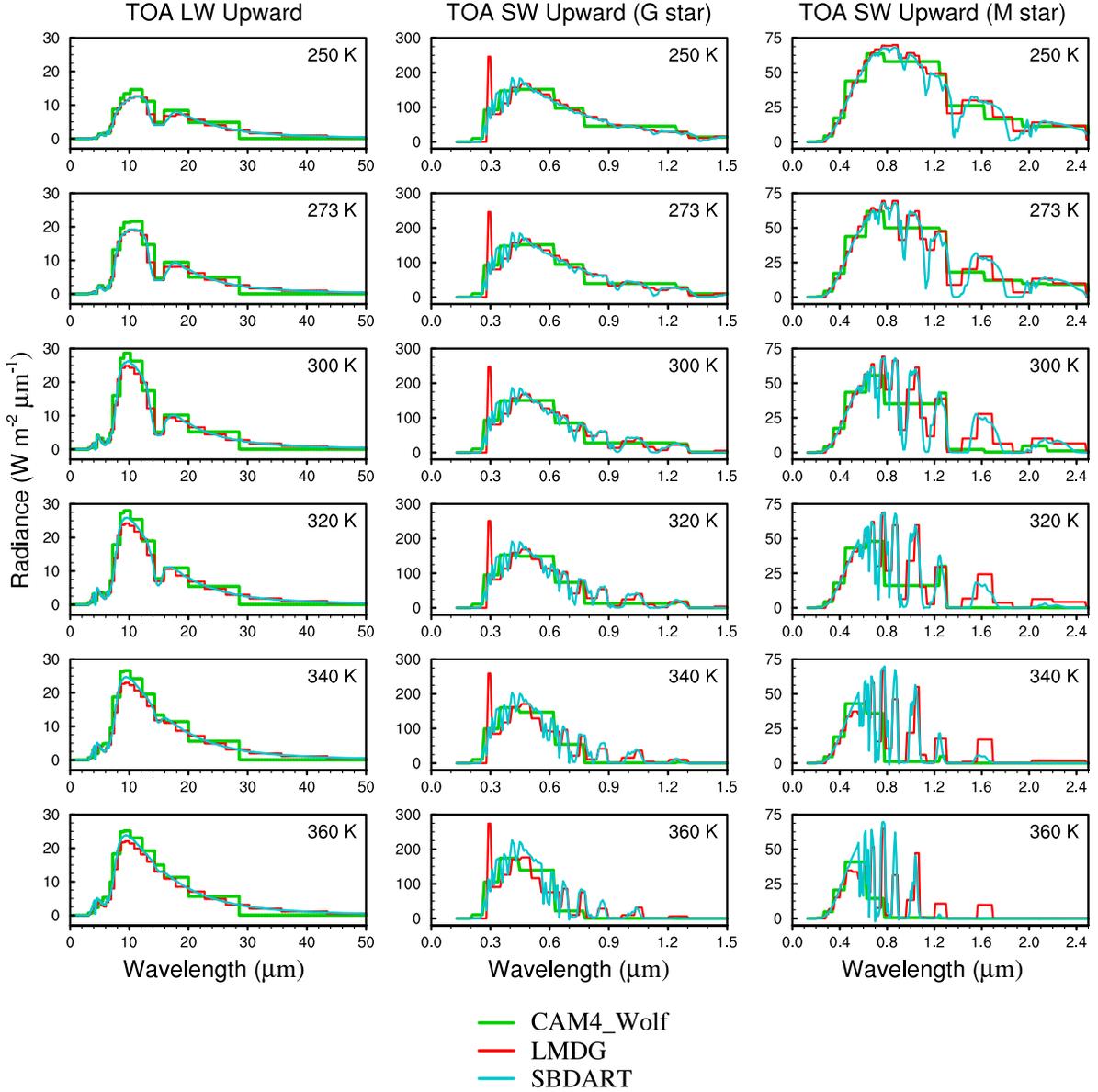}
\vspace{-15mm}
\caption{Longwave and shortwave spectra for the three correlated-$k$
  models, CAM4\_Wolf, LMDG, and SBDART. First
  column: upward longwave spectra at the top of atmosphere; second
  column: upward shortwave spectra at the top of atmosphere for G
  star; and third column: same as the second column, but for M star. In
  LMDG, a sharp bump exists at around 0.28 $\mu$m for G star (second
  column); this is because LMDG assigns all energy at wavelengths
  shorter than 0.28 $\mu$m to the narrow band between 0.28--0.30
  $\mu$m. Spectra from CAM3 and AM2 are unavailable.}
\label{fig-jun-LW-SW-spectra}
\end{center}
\end{figure}

Between the two line-by-line models, SMART and LBLRTM, the difference
in the OLR is mainly from the H$_2$O window region around 10~$\mu$m,
where SMART absorbs less energy than LBLRTM (Fig.~\ref{fig-colin-LW-spectra}). 
This is likely due to different assumptions in water vapor continuum absorption (Table~1). 
Among the three models using the correlated-$k$ method, LMDG, CAM4\_Wolf, and 
SBDART, the difference in the OLR is also mainly from the window region of around 
10 $\mu$m (left panels of Fig.~\ref{fig-jun-LW-SW-spectra}), which again emphasizes
that differences in water vapor continuum absorption assumptions likely drive the 
inter-model spread in longwave behavior. Additionally, LMDG and SBDART appear to 
emit more than CAM4\_Wolf at wavelengths longer than 28 $\mu$m.


\section{Comparison of Shortwave Radiation}\label{section-results-SW}

\begin{figure}
\begin{center}
\includegraphics[angle=0, width=165mm]{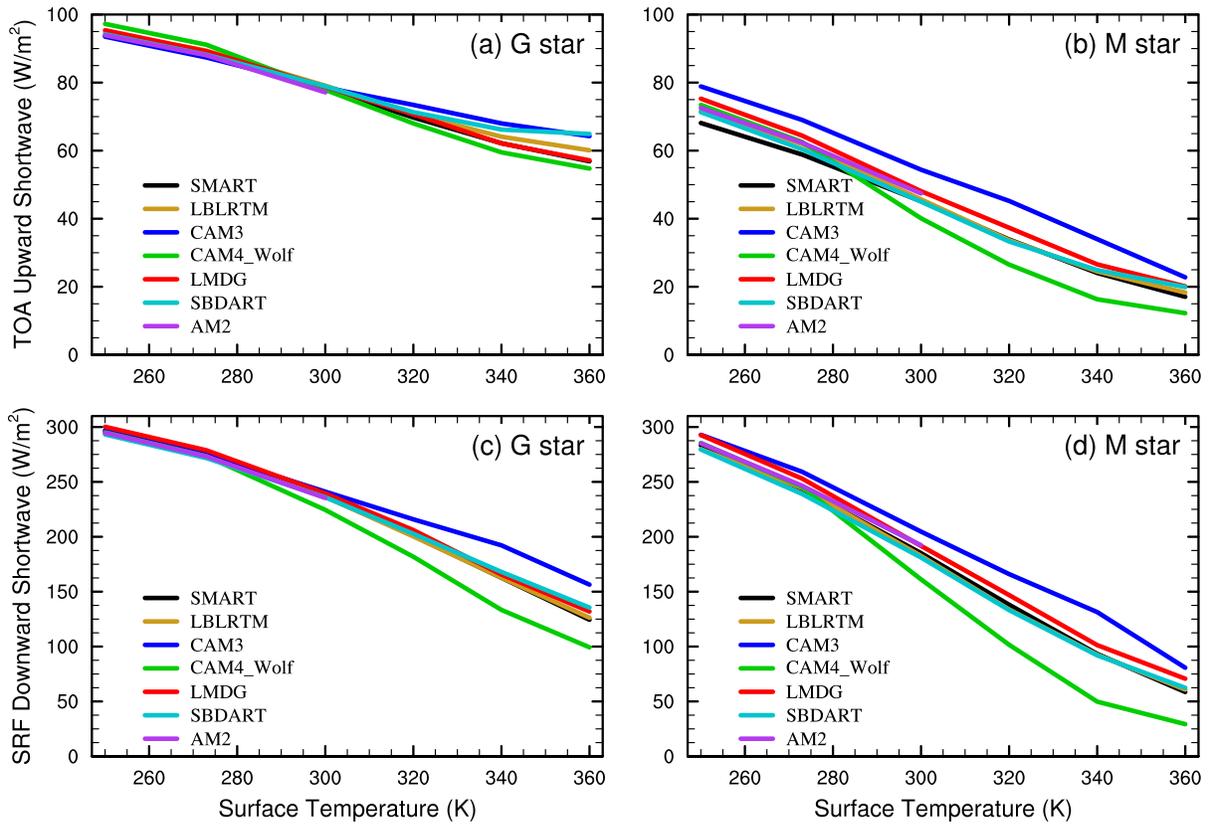}
\vspace{-30mm}
\caption{Shortwave fluxes for a G star (left panels) and for an M star
  (right panels) as a function of surface temperature from 250 to
  360~K for all models: SBDART, LMDG, AM2, CAM3, CAM4\_Wolf, SMART, and LBLRTM. 
  (a) and (b): upward shortwave flux at the top
  of atmosphere; (c) and (d): downward shortwave flux at the surface.}
\label{fig-ASR-1D}
\end{center}
\end{figure}

\begin{figure}
\begin{center}
\includegraphics[angle=0, width=165mm]{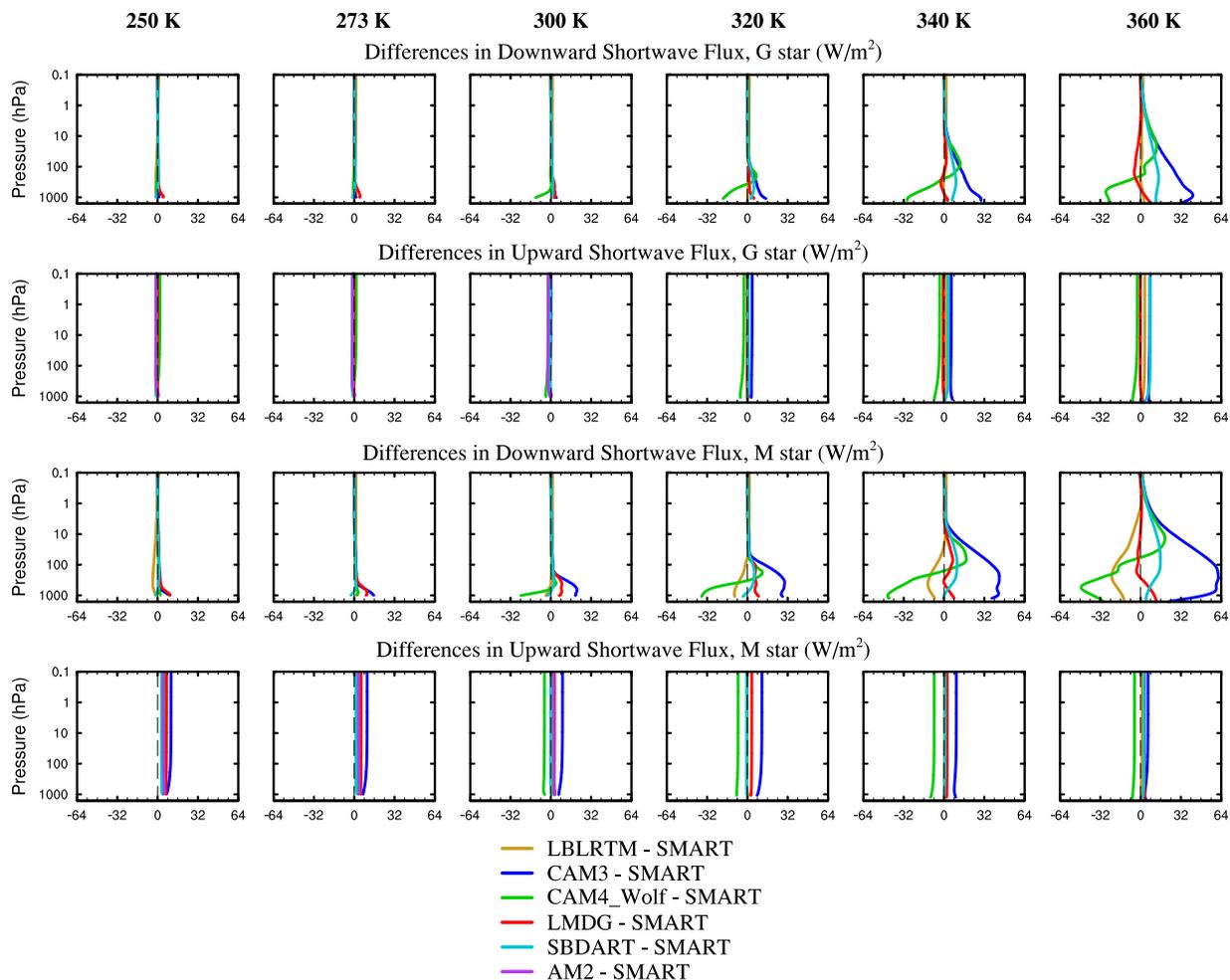}
\vspace{-15mm}
\caption{Differences in shortwave flux between LBLRTM, CAM3,
  CAM4\_Wolf, LMDG, SBDART, AM2, and SMART as a function of pressure
  and for surface temperatures from 250 to 360~K. First row: Downward
  shortwave flux for a G star; second row: upward shortwave flux for a
  G star; third row: downward shortwave flux for an M star; and fourth
  row: Upward shortwave flux for an M star. }
\label{fig-SW-flux-pressure}
\end{center}
\end{figure}

Fig.~\ref{fig-ASR-1D} shows upward shortwave radiation flux at the TOA
and downward shortwave radiation flux at the surface as a function of
surface temperature. All the models show that the upward and downward 
shortwave fluxes decrease with increasing surface temperature. This is mainly due to
the increase in shortwave absorption by water vapor. Under the solar
spectrum, the difference increases with temperature and the maximum
difference of the shortwave flux among the models is less than
$\approx$\,10~W\,m$^{-2}$ at the TOA, but can reach 60~W\,m$^{-2}$ at
the surface. The increase in spread of the surface flux among models
at higher temperatures is mostly due to the divergence in behavior of
only two models, CAM3 and CAM4\_Wolf. In general, the differences in
downward shortwave flux near the surface and in the troposphere are
much larger than those at the TOA (Fig.~\ref{fig-SW-flux-pressure}).
Moreover, the differences in upward shortwave flux among the models are 
much smaller than those in downward shortwave flux. This is due to the fact 
that the surface absorbs 75\% of the downward energy.

Under the M-star spectrum, the upward shortwave flux at the TOA and
the downward shortwave flux at the surface are much less than those
under the solar spectrum (right panels of Fig.~\ref{fig-ASR-1D}). This
is due to the redder  M-star spectrum, which leads to more
absorption by water vapor and less Rayleigh scattering
\citep{Pierrehumbert2010}. The maximum difference among the models,
however, is larger than that under the solar spectrum: 20~W\,m$^{-2}$
at the TOA and 90~W\,m$^{-2}$ at the surface. For both G- and M-star
spectra, CAM3 has the smallest shortwave absorption and the largest
Rayleigh scattering, while CAM4\_Wolf has the largest shortwave
absorption and the smallest Rayleigh scattering, especially once
the surface temperature is equal to or higher than 300~K. When the surface 
temperature is less than 300~K, the difference among all the models is
less than 15~W\,m$^{-2}$.

These models use different absorption line databases, different
spectral resolutions, different methods for H$_2$O continuum
absorption, and different multiple scattering schemes (see
Table~\ref{model-descriptions}), all of which affect the shortwave
radiation. For instance, as shown in the Supplementary Information of
\citet{Goldblattetal2013}, the new HITRAN line
  database (e.g., HITRAN2008) has stronger shortwave absorption than
the old HITRAN line database (e.g., HITRAN2000).
Moreover, CAM3, AM2, and CAM4\_Wolf have very coarse spectral
resolutions, while LBLRTM, SMART, SBDART and LMDG
have relatively fine resolutions. This could also cause differences
among the models. Preliminary tests using CAM4\_Wolf
  show that increasing the number of spectral intervals causes a
  significant improvement on the shortwave flux calculations
  (Kopparapu, Haqq-Misra, \& Wolf, in preparation). The high
  resolution codes---LBLRTM, SMART, and SBDART are able to resolve
  individual absorption bands of water vapor in the near-infrared
  region (Figs.~5 and 6). LMDG, which has 36 shortwave spectral
  intervals, is also able to somewhat resolve the individual
  absorption and window bands separately (Fig.~6). On the contrary,
  CAM4\_WOLF, CAM3, and AM2 must combine numerous near-infrared bands
  and window regions into larger spectral intervals. This can lead to
errors, such as the fact that near-infrared absorption by CO$_2$ at
4.3~$\mu$m is not considered in CAM3 \citep{Collinsetal2006}.
There may also be some unknown parameters and errors in individual
models, which could cause differences among the models.

\begin{figure}
\begin{center}
\includegraphics[angle=0, width=165mm]{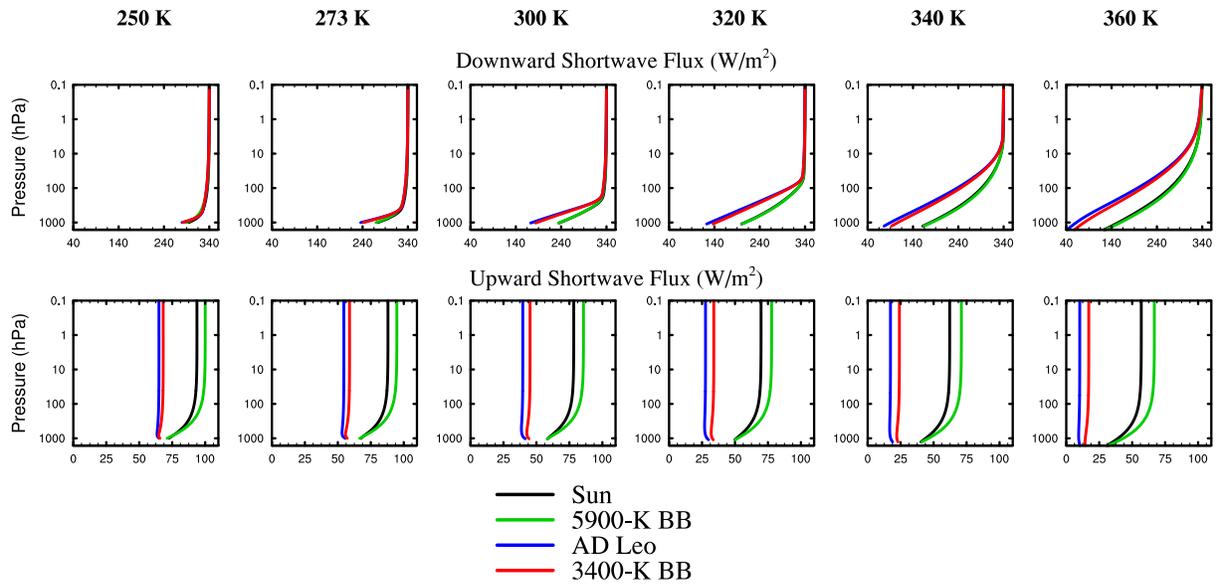}
\vspace{-50mm}
\caption{Shortwave fluxes for four different stellar spectra: the Sun,
  a 5900-K blackbody, AD Leo, and a 3400-K blackbody, calculated using
  the line-by-line model SMART. Upper panels: downward shortwave flux;
  lower panels: upward shortwave flux.}
\label{fig-SMART-4spectrums}
\end{center}
\end{figure}

The influence of stellar spectrum on the radiative fluxes is further
shown in Fig.~\ref{fig-SMART-4spectrums}. Using the line-by-line model
SMART, we calculate downward and upward shortwave radiative fluxes
under four different stellar spectra: the Sun, a 5900-K blackbody, the AD Leo,
and a 3400-K blackbody. Primarily due to the wavelength dependence of
shortwave absorption by water vapor, the differences in the radiation
fluxes are significant between G- and M-star spectra, as has been previously
found by others \citep[e.g.,][]{Kastingetal1993, Pierrehumbert2010, 
Kopparapuetal2013, Shieldsetal2013, Shieldsetal2014, Godoltetal2015}.
Rayleigh scattering is also wavelength-dependent, but its effect is
much smaller than that of water vapor absorption
\citep{Halthoreetal2005}. Moreover, the differences between a
realistic stellar spectrum and its corresponding blackbody spectrum
are relatively small, for both G and M stars. For instance, the
maximum difference in the upward shortwave flux at the TOA between the
AD Leo and the 3400-K blackbody is only $\approx$\,7~W\,m$^{-2}$.


\section{Discussion: The Inner Edge of the Habitable Zone}

\begin{figure}
\begin{center}
\includegraphics[angle=0, width=165mm]{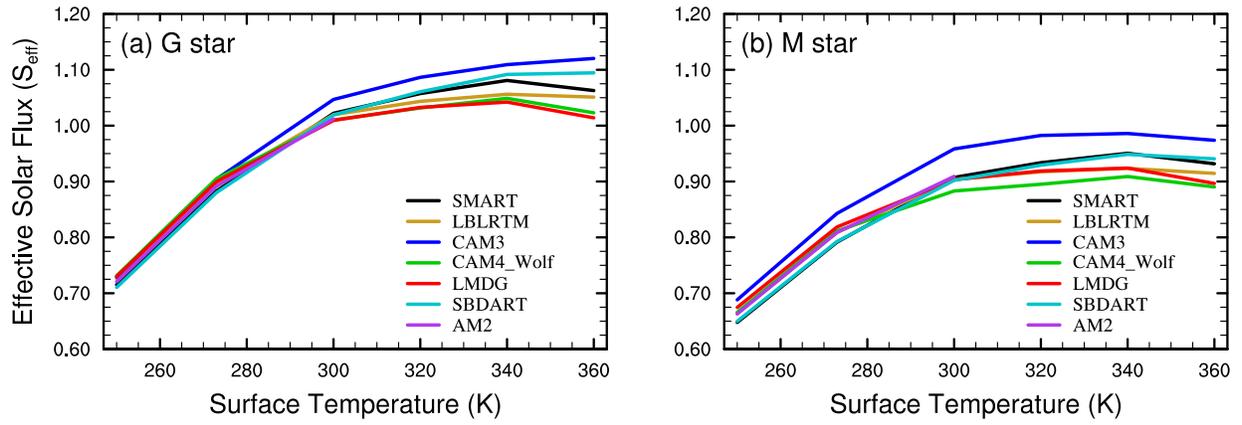}
\vspace{-60mm}
\caption{Effective solar flux for all models for (a) a G star and (b)
  an M star. The effective solar flux is defined as the ratio of
  outgoing longwave radiation to absorbed shortwave radiation. It is
  the factor by which present Earth's solar flux would need to be
  multiplied in order to maintain a given surface temperature.}
\label{fig-effective-solar-flux}
\end{center}
\end{figure}

Now that we have explored the differences in radiative transfer among
the models, it is important to put these differences in context in
terms of the effect they can have on the inner edge of the habitable
zone. One way we can approximate the inner edge based on our fixed
atmospheric temperature profile simulations is by using the effective
solar flux \citep[S$_\textrm{eff}$,][]{Kasting1988}, which is defined as the
ratio of the outgoing longwave radiation to the total shortwave
radiation absorbed by the planet (net shortwave at the TOA). S$_\textrm{eff}$ corresponds 
to the factor by which one would have to multiply Earth's solar flux in order to 
maintain a given surface temperature, and we have plotted it in
Fig.~\ref{fig-effective-solar-flux}. The maximum difference in
S$_\textrm{eff}$ is $\approx$\,3\,\% between the two line-by-line models and
$\approx$\,10\,\% among the band models, which corresponds to $\approx$\,10
and $\approx$\,34 W\,m$^{-2}$ in the global mean, respectively. This shows
that  uncertainty in radiative transfer, even neglecting more
complicated processes such as clouds and areas of sub-saturation, has a
fairly significant effect on estimates of the inner edge of the
habitable zone.

We should also note that similarity in S$_\textrm{eff}$ can mask differences
in model behavior. For example CAM4\_WOLF and LMDG produce very
similar values of S$_\textrm{eff}$ (Fig.~\ref{fig-effective-solar-flux}), but
this results from the fact that, at a given surface temperature,
CAM4\_WOLF both emits more outgoing longwave radiation
(Fig.~\ref{fig-OLR-1D}) and absorbs more shortwave
(Fig.~\ref{fig-ASR-1D}) than LMDG. Moreover, even though the 1D
calculations indicate that CAM4\_WOLF and LMDG produce a runaway
greenhouse at a stellar flux within about 1\,\% of each other for a
G-star spectrum, 3D calculations show that a runaway greenhouse occurs
for Earth when the solar constant is increased by 10\,\% in LMDG
\citep{Leconteetal2013}, and has not yet occurred when the solar
constant is increased by 21\,\% in CAM4\_WOLF \citep{WolfandToon2015}.
This difference could be caused simply by differences in simulations of
sub-saturated regions and clouds, or it could result from more
complicated feedbacks between atmospheric dynamics and the detailed
differences in longwave and shortwave behavior between the two models
mentioned above. This motivates a full comparison of 3D global climate
models, which we are currently pursuing.

It would be encouraging if we could find some sort of relation between
the age of the line database and water vapor continuum assumptions
made by the models (Table~\ref{model-descriptions}) and similarity in
model behavior. In the longwave, this is certainly not the case. For
example, the models with the oldest databases (SBDART 
with HITRAN1996) and the newest databases (SMART 
with HITEMP2010 and HITRAN2012) yield almost identical outgoing longwave 
radiation (Fig.~\ref{fig-OLR-1D}). In the shortwave, it does appear that models
using databases developed within the past ten years clump together
(Fig.~\ref{fig-ASR-1D}), which may indicate that our understanding of
shortwave absorption by large amounts of water vapor is converging,
although it is always possible that the next generation databases will
overturn this trend. In any case, our work emphasizes the need to
develop more accurate line and continuum databases, and to try to
constrain them with actual data (rather than just theoretical
calculations) insofar as this is possible.

\section{Conclusion}

We have compared seven radiative transfer models that are currently
being used to estimate the inner edge of the habitable zone. We found
that there are significant differences among the models in both
shortwave and longwave radiative fluxes, especially in the troposphere
and at the surface. The maximum difference in radiative fluxes is on
the order of tens of watts per square meter, and the uncertainty in
estimating the insolation threshold of the inner edge of the habitable
zone is about 10\,\% of the present Earth's solar constant
among the band models and about 3\,\% between the two line-by-line
  models. Our results suggest two ways to improve the radiative
transfer in climate simulations of exoplanets. One is to improve the
absorption coefficients and continuum behavior of water vapor,
especially in the infrared window region and in the entire visible
region, and the other one is to increase the resolution of stellar
spectra in broadband models.

\acknowledgments \textbf{Acknowledgments:} We are grateful to Robin
Wordsworth for insightful discussions, and to Jonah Bloch-Johnson and
Xiaoxiao Tan for their help in radiative transfer calculations. We
thank Rodrigo Caballero for maintaining CliMT, which we used in the
project.

Software: SBDART, AM2 (1D), CAM3 (1D), CAM4\_Wolf (1D), LMDG (1D),  SMART, LBLRTM




\newpage
\clearpage
\begin{table}[h]\footnotesize
\begin{center}
  \caption{Main characteristics of 1D radiative transfer models$^a$ employed in our intercomparison, 
  including spectral line databases for H$_2$O absorption (HITRAN), the number of vertical levels (Lev.),
    the number of intervals for stellar spectra (Int.), and methods for
    calculating absorption coefficients, H$_2$O continuum absorption (H$_2$O Cont.),
    and multiple scattering. \label{model-descriptions}}
\begin{tabular}{llllllll}
  \tableline\tableline\noalign{\smallskip}
  Models$^b$ & HITRAN & Lev. & Int. & Absor. Coeff. & H$_2$O Cont. & Multiple Scattering  & Examiners  \\
  \tableline\noalign{\smallskip}
SBDART & 1996     & 301      & 369           & correlated-$k$    & CKD2.3          &   \citet{Stamnesetal1988}           & Wang\\
CAM3$^c$  & 2000    & 301    & 19           & absorp./emis.     & CKD2.4          &  \citet{Briegleb1992}                  & Yang\\
AM2        & 2000        & 301    & 18         & exponential sum      & CKD2.1          &  \citet{EdwardsandSlingo1996}  & Feldl \\
CAM4\_Wolf &  2004       & 301  & 23       & correlated-$k$   & MT\_CKD2.5  & \citet{Toonetal1989}                 & Wolf\\
LMDG  & 2008  & 301              & 36          & correlated-$k$   & CKD2.4           & \citet{Toonetal1989}                 & Leconte\\
LBLRTM & 2008  & 150$^d$   & $>$$10^{4}$                  & line-by-line       & MT\_CKD2.5   & \citet{MoncetandClough1997} & Wolf\\
SMART & 2010$^e$ & 75      & $>$$10^{4}$            & line-by-line        & $\chi$-factors    & \citet{Stamnesetal1988}            & Goldblatt\\

  \noalign{\smallskip}\tableline
  \end{tabular}
\end{center}
\emph{a.}~SBDART is a software tool for computing radiative transfer, 
developed by \citet{Ricchiazzietal1988}. CAM version 3 (CAM3) is a 3D atmospheric general 
circulation model (GCM), developed at NCAR. 
CAM4\_Wolf is CAM version 4 (CAM4) but with a new radiative transfer module, developed 
by E.~T.~Wolf. AM2 is a 3D GCM developed at NOAA/GFDL. 
LMDG is the 3D Laboratoire de M\'et\'eorologie Dynamique (LMD) Generic Model, 
developed at LMD, Paris, France. Here, CAM3, AM2, CAM4\_Wolf, and LMDG are the pure radiative transfer 
modules of the corresponding 3D GCMs. SMART is a line-by-line radiative transfer models, developed 
by David Crisp at NASA's JPL in California. LBLRTM is another line-by-line model developed 
at the Atmospheric and Environmental Research, Inc. (AER).\\
\emph{b.}~Appropriate references for SBDART: \citet{Ricchiazzietal1988} and \citet{Yangetal2000}; 
for CAM3: \citet{Collinsetal2002} and \citet{RamanathanandDowney1986}; 
for AM2: \citet{EdwardsandSlingo1996} and \citet{FreidenreichandRamaswamy1999}; 
for CAM4\_Wolf: \citet{WolfandToon2015};  
for LMDG: \citet{Wordsworthetal2010a, Wordsworthetal2010b}; 
for LBLRTM:  \citet{Cloughetal2005, Cloughetal1992};  
and for SMART: \citet{MeadowsandCrisp1996} and \citet{Crisp1997}. 
CAM3 uses an absorptivity/emissivity formulation for absorption 
coefficients. Appropriate references for H$_2$O continuum absorption are 
\citet{Cloughetal1989, Cloughetal2005} and \citet{Mlaweretal2012}. 
All the line databases are developed at the Atomic and Molecular Physics Division, 
Harvard-Smithsonian Center for Astrophysics under the direction of L.\,S.~Rothman 
\citep[see ][ and references therein]{Rothmanetal2013}. \\
\emph{c.}~Pure radiative transfer calculations with CAM3 were done with CliMT (http://climdyn.misu.su.se/climt/). 
CliMT is an object-oriented climate modeling and diagnostics toolkit, developed by Rodrigo Caballero. \\
\emph{d.}~150 levels for longwave calculations, and 75 levels for shortwave calculations. \\
\emph{e.}~HITEMP2010 for H$_2$O and HITRAN2012 for CO$_2$. \\  
  
\end{table}


\clearpage 

\clearpage 

\clearpage 

\clearpage 

\clearpage 

\clearpage 

\clearpage 

\clearpage 

\clearpage 

\clearpage 

\end{document}